\documentclass[preprint,12pt,authoryear]{elsarticle}

\usepackage{amssymb}
\newcommand{\cmark}{\ding{51}}%
\newcommand{\xmark}{\ding{55}}%
\usepackage{graphicx}
\usepackage{epstopdf}
\usepackage{alltt}
\usepackage{multirow}
\usepackage{makecell}
\usepackage{hhline}
\usepackage{lipsum}
\usepackage{float}
\usepackage{amssymb}
\usepackage{pifont}
\usepackage{stfloats}
\usepackage{textcomp}
\usepackage{xcolor}
\usepackage{booktabs, caption, makecell}

\usepackage{threeparttable}

\journal{Computer Networks}

\begin{document}

\begin{frontmatter}





\title{Cyber Security Challenges and Solutions for V2X Communications: A Survey}

\author[mymainaddress,mysecondaryaddress]{Aljawharah Alnasser}
\ead{alalnasser@ksu.edu.sa}

\author[mysecondaryaddress]{Hongjian Sun\corref{mycorrespondingauthor}}
\cortext[mycorrespondingauthor]{Corresponding author}
\ead{hongjian.sun@durham.ac.uk}

\author[mythirdadress]{Jing Jiang}
\ead{jing.jiang@northumbria.ac.uk}

\address[mymainaddress]{Department of Information Technology, King Saud University, Riyadh, Saudi Arabia}
\address[mysecondaryaddress]{Durham University, Durham, UK}
\address[mythirdadress]{Department of Mathematics, Physics and Electrical Engineering, Northumbria University, UK}

%
%
%

\begin{abstract}
	In recent years, vehicles became able to establish connections with other vehicles and infrastructure units that are located in the roadside. In the near future, the vehicular network will be expanded to include the communication between vehicles and any smart devices in the roadside which is called Vehicle-to-Everything (V2X) communication. The vehicular network causes many challenges due to heterogeneous nodes, various speeds and intermittent connection, where traditional security methods are not always efficacious. As a result, an extensive variety of research works has been done on optimizing security solutions whilst considering network requirements. In this paper, we present a comprehensive survey and taxonomy of the existing security solutions for V2X communication technology. Then, we provide discussions and comparisons with regard to some pertinent criteria.
Also, we present a threat analysis for V2X enabling technologies. Finally, we point out the research challenges and some future directions.

\end{abstract}


\end{frontmatter}
\section{Introduction}
\label{}
As a result of massive spread of Internet, a new notion has emerged which converts rigid objects to smart objects and connects them together, known as Internet of Things (IoT). This is achieved by embedding extra hardware 
\begin{table}[t!]
	\caption {Abbreviations used throughout the paper.} \label{tab:title} 
	\begin{center}
		\scalebox{0.7}{\begin{tabular}{ | l | c |}
			\hline
		3GPP &  $3^{rd}$ Generation Partnership Project \\ \hline
			D2D & Direct-to-Direct \\ \hline
			eNodeB & Evolved Node B  \\ \hline
			IoT & Internet of Things \\ \hline
			LTE & Long-Term Evolution \\ \hline
			RSU & Road Side Unit \\ \hline
			UE & User Equipment \\ \hline
			V2I & Vehicle-to-Infrastructure \\ \hline
			V2P & Vehicle-to-Pedestrian \\ \hline
			V2V & Vehicle-to-Vehicle \\ \hline
			V2X & Vehicle-to-Everything \\ \hline
        	QoS & Quality of Service \\ \hline
			
		\end{tabular}}
	\end{center}
\end{table}
such as sensors and communication interface within each device and combining them with a software system. Thus, the devices can sense the surrounding environment and share information using wireless communications. 

IoT has been broadly applied in various domains such as health-care, smart cities and industry. Indeed,  the people spend the most times in homes, offices and transportation. According to the U.S. Department of Transportation and Safety Administration, the people spend 500 million hours per week in the vehicle. As a result, based on Alcatel-Lucent's research which was accomplished in 2009, they found that over 50\% of participants liked the idea of connected vehicles and 22\% are willing to pay fees for communication services (\cite{general18}).	
Consequently, the need has been increased for joining transportation system under the umbrella of IoT. Initially, transportation system was transformed into cyber physical system by embedding software into vehicles. Hence, vehicles can form a network and communicate with any smart device as a part of IoT.

With the integration of IoT technologies, vehicular networks became vulnerable to various types of cyber-attacks: internal or external attacks. Internal attacks are initiated by the fully authorized node which can bypass the authentication model, while external attacks are launched by an unauthorized node. In the latter case, secure authorization model can minimize the effect of these attacks. 

Many security solutions were studied to protect the vehicular networks against various types of cyber attacks. In this survey, we aim to provide a deep research in this area and study the security solutions for all types of vehicular communications. We believe that this survey will guide future research for addressing challenges of the future vehicular network.
\subsection{Relations to Existing Surveys}
Although there are a large number of publications regarding the security aspects in vehicular networks, there is so far no comprehensive survey on that for Vehicle-to-Everything (V2X) communications. Vehicular ad-hoc network is one type of vehicular networks. There exist surveys focusing on general structure of vehicular ad-hoc network: \cite{general13} provided a survey on vehicular ad-hoc network structure, enabling technologies, applications, and research challenges. \cite{general14} proposed the basic characteristics and requirements of vehicular ad-hoc network, and the standardization efforts in intelligent transportation systems.

There are some surveys on general security challenges in vehicular ad-hoc networks only. For instance, \cite{general8} published a survey on the potential cyber attacks on vehicular ad-hoc networks and the proposed security methods to guard against them. \cite{general9} provided security requirements and threats in vehicular ad-hoc networks. Also, they discussed the attacks characteristics and security solutions. In addition, \cite{general10} reviewed the existing security solutions and described them on a comparable level. \cite{general12} presented a general review of some security research in vehicular ad-hoc network. 

On the other hand, there exist some surveys on a specific security mechanism that can be used in vehicular networks. \cite{general1} classified the security solutions in vehicular ad-hoc networks based on cryptographic schemes and compared them to evaluate their performance. While, \cite{general11} analysed existing trust-based solutions in multi-agent systems, mobile ad-hoc networks and vehicular ad-hoc networks. \cite{general2} reviewed the main existing trust-based models and studied when trust-based solution is more suitable than cryptography, and the opposite.
In this survey, we conduct an in-depth analysis on two dimensions. The first dimension includes the ability of security services in IEEE802.11p and Long-Term Evolution for V2X (LTE-V2X) to defend against various cyber-attacks. The second dimension covers the security challenges for various security methods which were applied on vehicular networks. This will five the researchers the full picture regards the security constraints for each communication protocol and also for each security method.

	\subsection{Contributions and Structure}
The main aim of this survey is to present a comprehensive and organized overview of various security solutions in vehicular networks and discuss the design challenges of security model for V2X communications. This paper makes three significant contributions to the field of V2X security:
\begin{enumerate}
	\item The proposed taxonomy classifies the security solutions of vehicular networks to study the challenges of designing security model for the V2X network which is a novel approach to the subject. 
	\item Threats analysis for V2X enabling technologies is conducted. Based on our knowledge, this is the first such analysis for threats in IEEE802.11p and LTE-V2X.
	\item We evaluate the effectiveness of security solutions on the considered attack, message type, latency limit and model structure.
	
\end{enumerate}

The paper is organized as follow: in section 2 we present some necessary background information.
In section 3 we mention the security requirements and threats analysis for V2X enabling technologies. 
A taxonomy of security methods for V2X technology is presented in Section 4. In Section 5, we classify and examine the presented methods. Some challenges and research direction are given
in Section 6. Finally, Section 7 concludes this survey.

\section{Vehicle-to-Everything (V2X) communication Technology}
V2X technology refers to intelligent transportation system where all road entities including vehicles, pedestrians, cycles, motorcycles and infrastructure units are interconnected with each other. This connectivity will produce more accurate information about the traffic situation across the entire network. Thus, it will help in improving traffic flows and reduce accidents. In 2015, Siemens implemented the first fully dynamic system on Germany's A9 highway. The result showed 35\% fewer accidents and reduction of people injured at roads with 31\% (\cite{t1}). 

\subsection{Architecture}
Intelligent transportation system applies data processing, communication, and sensor technologies to vehicles, infrastructure units and roadside users to increase safety and efficiency of the transportation system. The heterogeneous network consists of two main sub-networks (\cite{general15}) as shown in Figure 1:

\begin{figure}[b!]
	\includegraphics[width=\linewidth]{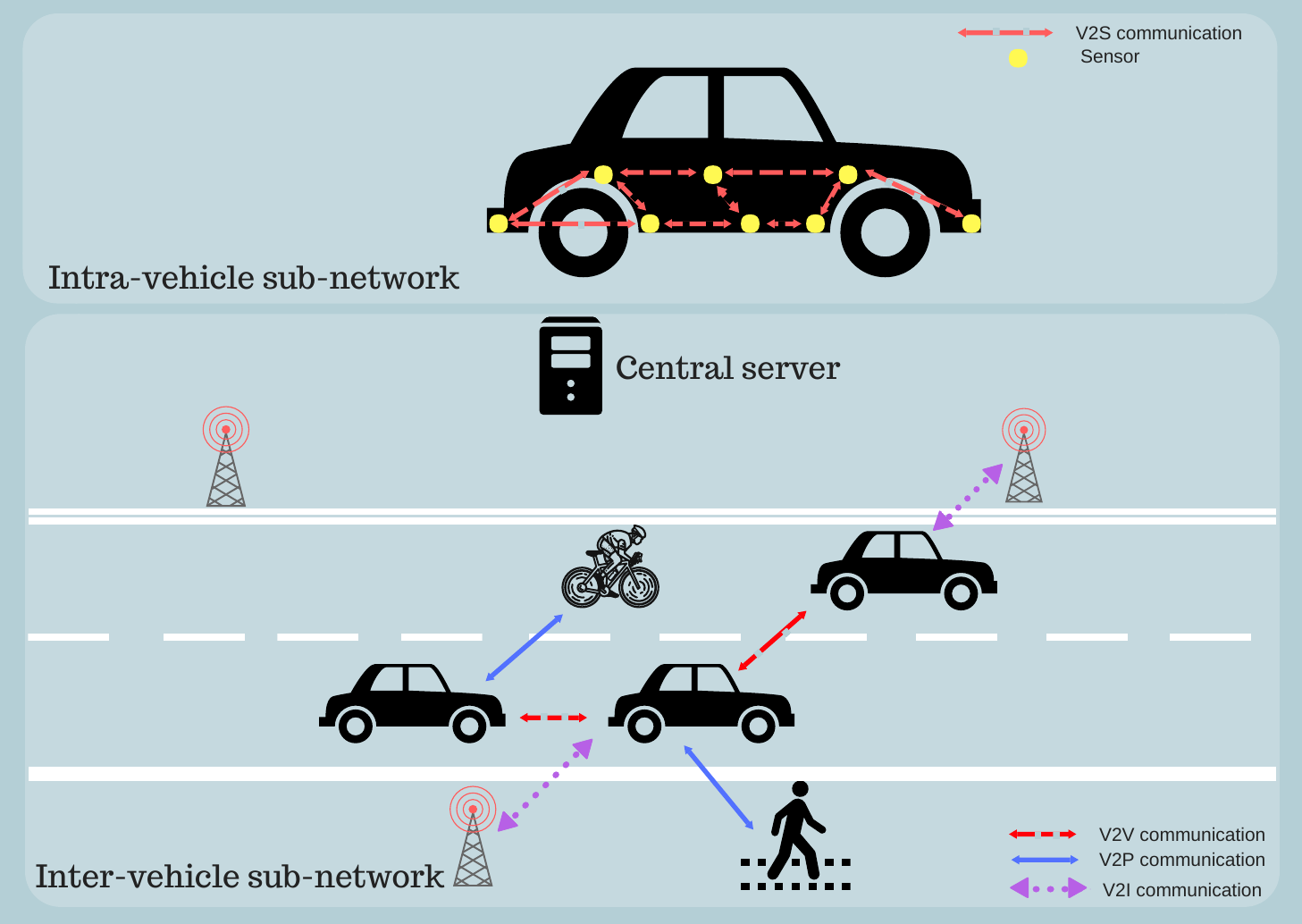}
	\caption{General Structure of intelligent transportation system.}
	
\end{figure}

	\begin{itemize}
	\item \textit{Intra-vehicle network} comprises of a collection of sensors which are located in the vehicle. The interactions among sensors are bridged via Ethernet, ZigBee or WiFi connections.
	\item \textit{Inter-vehicle network} covers the communication between the vehicle and surrounding devices. It comprises of four entities as follows:

	\begin{itemize}
		\item \textbf{On-board unit} is the main entity in intelligent transportation system. Each vehicle is equipped with on-board unit to be able to process the collected data and interact with surrounding entities.
		\item \textbf{Roadside users} such as pedestrian, motorcyclists, bikers and roller skates.
		\item \textbf{Road Side Unit (RSU)} is the transportation infrastructure unit which exist in the roadside. it has information about local topology which assists in providing several services to the road entities.
		\item \textbf{Central/Cloud server} has a central control on all road entities, traffic and roads.
	\end{itemize}
	
\end{itemize}

	\subsection{Communications}
V2X supports a unified connectivity platform for the connected entities. Also, it allows road entities to transmit information such as their current speed, position, and direction to their fixed and moving neighboring entities. Then, they use this information to make intelligent decisions. The communication type depends on the entities that establish the link. It supports five types of communications (\cite{general4}):
\begin{itemize}
	\item  \textit{Vehicle-to-Sensors (V2S)} represents the communication between sensors in intra-vehicle sub-network;
	\item \textit{Vehicle-to-Vehicle (V2V)} covers the communication between vehicles using V2V application;
	\item \textit{Vehicle-to-Pedestrian (V2P)} provides the connection between the vehicle and roadside users
	using V2V application;
	\item \textit{Vehicle-to-Grid (V2G)} supports the communication between vehicles and the electric grid to charge Electric Vehicles.
	\item \textit{Vehicle-to-Infrastructure (V2I)} represents the communication between road entities and infrastructure units.
\end{itemize}

In one vehicular network, all road entities are supposed to generate and exchange messages. The messages can be used to support variety of applications, e.g., applications related to safety, traffic and infotainment. The messages are categorized into four types (\cite{general5}):

\begin{itemize}
	\item \textbf{Periodic message (beacon):} Road entity periodically broadcasts a status message, which contains information such as speed, location and direction, to the neighboring entities. It generated at regular intervals between $100ms$ to $1s$. As a result, each entity can perceive the local topology. Also, they can predict and anticipate dangerous situations or traffic congestion. This type of messages is not time critical ($300ms$).

	\item \textbf{Local event triggered message:} Road entity sends the message when a local event is detected such as the critical warning or intersection assist. It is sent locally to the neighboring entities using V2V/V2P links where it contains useful information for neighborhood area only. In addition, it is a time critical which requires to be delivered with a low latency around $100ms$.
	%
	\item \textbf{Global event triggered message:}  Road entity sends the message when a global event is detected such as road construction and road congestion. This message needs to be propagated over a wider area. As a result, road entities use V2I communication link to transmit the message. 
	
	\item \textbf{Emergency vehicle message:} It is used to support a smooth movement for emergency vehicles. It is sent by emergency vehicles to the surrounding vehicles using V2V/V2P links to clear the road.
	
\end{itemize}

	\subsection{Applications}

As a result of the technological improvements in the areas of sensing and wireless networking, intelligent transportation system permits for the existing of various applications that are related to safety, traffic, and infotainment (\cite{general15}). 
\begin{itemize}
	\item \textit{Safety-related applications} use wireless communications between surrounding entities to decrease accidents and protect the commuters from dangers. Each road entity periodically sends safety message to its neighbors to report its current status. Furthermore, they may also need to transmit warning messages when local or global event is detected.
	\item \textit{Traffic-related applications} are deployed to manage the traffic efficiently and ensure smoothly traffic flow. They are responsible for collecting the traffic information and transmitting them wirelessly to a remote server for analysis. After that, the analysis results are sent to vehicles for future usage.
	
	\item \textit{Infotainment-related applications} aim at improving the driving experience by supporting various services such as Internet access, online gaming, video streaming, weather information.
	
\end{itemize}

	\section{Security for V2X Enabling Technologies}
Supporting safety-related applications is the core of vehicle-to-vehicle communication. Since ten years ago, V2X technology has been enabled by IEEE802.11p, which has been standardized, implemented and examined. 

One of the most critical challenges to make V2X technology feasible is how to ensure interoperability among heterogeneous devices. As a result, the $3^{rd}$ Generation Partnership Project (3GPP) has worked on standardization for LTE protocol to fit the requirements and services of the V2X communications. 3GPP has concentrated on supporting different types of communications using one standard. The first release (Release 8) was in 2008.  The standardization of LTE Advanced Pro (Release 14) finalized at the beginning of 2017 (\cite{general4}).

 The safety of commuters relies on the performance of these technologies. Consequently, it is important to analyse the security services offered by these technologies.
\subsection{IEEE 802.11p}
\subsubsection{Overview}
It is an enhanced version of the ad-hoc mode in IEEE802.11a. It was implemented for supporting the communication between mobile nodes with the presence of obstacles, dynamic topology and intermittent connection. The main purpose of IEEE802.11p is to support non-line of sight (\cite{general17}).
It was proposed for supporting intelligent transportation system applications in vehicular adhoc networks. 
It provides ad-hoc communication between vehicles and RSUs. It gives vehicles the ability to share information with their neighboring vehicles using V2V and V2I only. 


IEEE 802.11p can be easily deployed with minimum cost, however, it lacks of scalability, unlimited delays, and Quality of Service (QoS). Furthermore, it can only offer intermittent V2I connectivity because of the short radio range (\cite{ref23}). 

	\subsubsection{Security Measures}
IEEE P1609.2 is based on cryptography standards such as elliptic curve cryptography, wireless access in vehicular environment certificate formats, and hybrid encryption methods (\cite{ref16}).
\textit{Broadcast Messages} usually are not directed to particular destination and they related to safety-related applications. Also, they contain the timestamp which is obtained from the internal clock for synchronization. However, these messages are only signed with the sender's certificate. Elliptic Curve Digital Signature Algorithm is the signature algorithm in the standard (\cite{ref15}). 
\textit{Transaction Messages} are generally unicast messages and they may be used to access location-based services and personal data. Consequently, to protect the data, these messages are encrypted with a symmetric encryption algorithm. To ensure more protection, the algorithm uses random key which is encrypted using elliptic curve integrated encryption scheme (\cite{ref15}).

\subsection{LTE-V2X}

	\subsubsection{Overview}
It has the potential to deal with the low-latency and high-reliability V2X use cases. LTE-V2X is mainly composed of six main components (\cite{general4}) as shown in Figure 2:

\begin{itemize}
	\item \textit{User Equipment (UE)} is the device that is used directly by an end-user to communicate with eNodeB or other UEs.
	\item \textit{Evolved Node B (eNB)} is the wireless interface for LTE network which allows for sending and receiving radio transmissions to/from all UEs in one or more cells. 
	\item \textit{V2X Application Server} is responsible for distribution of V2X messages to different target areas.
	\item \textit{V2X Control Function} is responsible for authorization and revocation of V2X services. It provides various services after successful mutual authentication and security key generation.
	\item \textit{Multimedia Broadcast Multicast Service} supports efficient delivery for multicast services over areas typically spanning multiple cells.
	\item \textit{Single-cell Point-to-Multipoint} provides the delivery of multicast services over a single cell.
	
\end{itemize} 
	\begin{figure}[t!]
	\includegraphics[width=\linewidth]{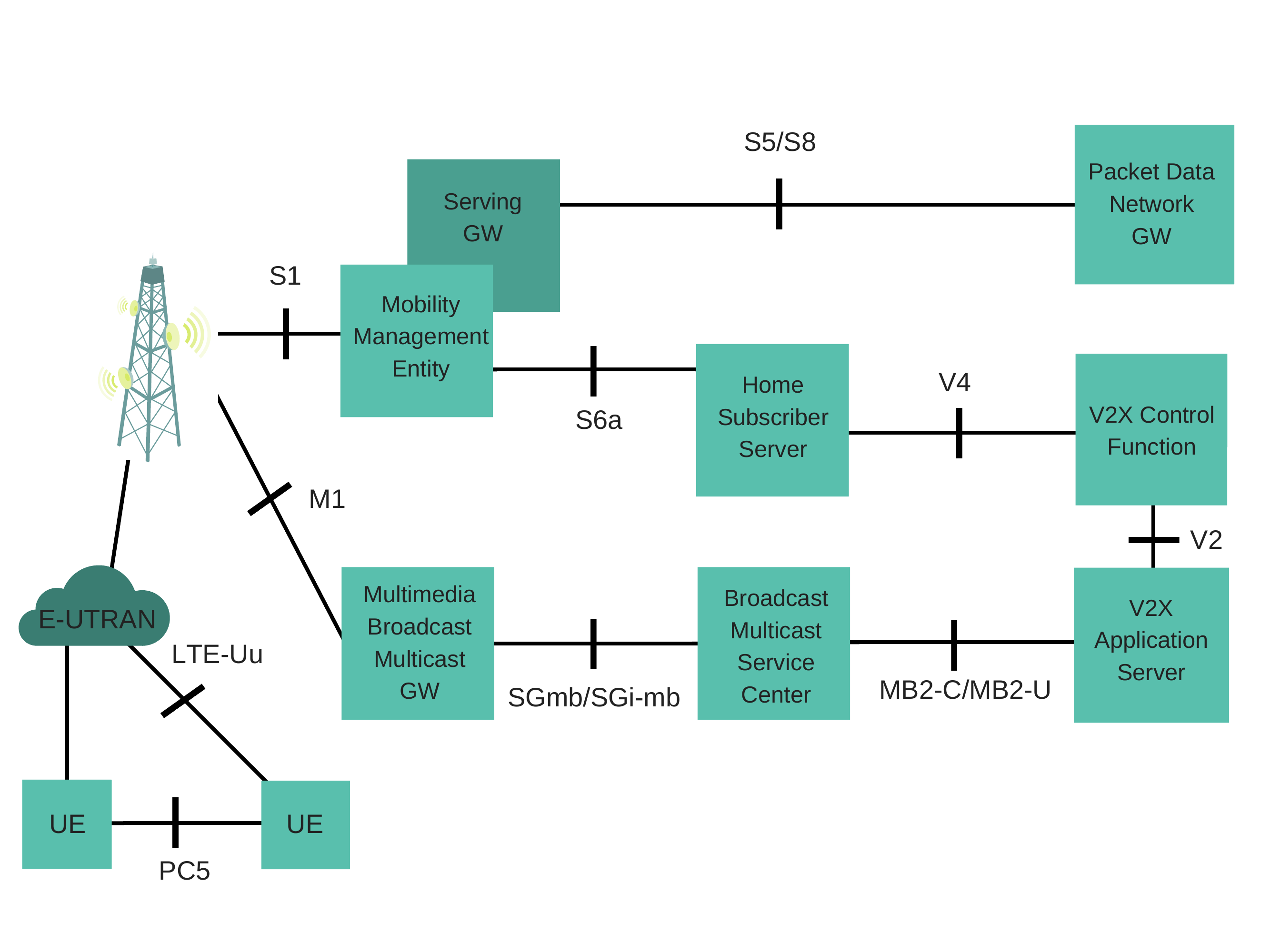}
	\caption{General Structure of LTE-based V2X network.}
\end{figure}

In LTE-V2X, the messages are sent using two types of links, as shown in Figure 3:

\begin{itemize}
	\item \textit{Cellular-based communication} covers the two way communication between UE and eNB over LTE air interface (\cite{general16}). The communication going from UE to eNB is called uplink and when it is going from eNB to a UE it is called downlink. Cellular-based communication covers wide area with high capacity. It is used by V2X application server to broadcast messages to vehicles and beyond, or send them to the server via a unicast connection. In addition to one-to-one communications between eNB and UE, eNB supports one-to-many communications via downlink. eNB uses single-cell point-to-multipoint service for the transmissions over a single cell and multimedia broadcast multicast service for communications over multiple cells.
	
	\item \textit{Device-to-Device communication (D2D)} enables the direct connection between UEs without traversing eNB and it is called side-link. It supports multi-hop communications between network entities to enhance the end-to end connectivity. Also, it provides a short-range communication and low latency for safety messages. It allows for UE to transmit data directly to other UEs over the side-link even if they reside out-of-network coverage. Every D2D pair can communicate via $Inband$ or $Outband$ modes (\cite{general16}).
	\textbf{Inband mode}  uses the cellular spectrum for both D2D and cellular communications. \textit{Underlay communication} allows for both of them sharing and reusing the same radio resources to improve the spectrum efficiency. The main drawback is the high possibility of collision between D2D links and cellular links.
	In contrast, \textit{overlay communication} allocates dedicated cellular resources for D2D connections between the transmitter and the receiver. To minimize the interference between D2D and cellular links, \textbf{outband mode} uses unlicensed spectrum such as 2.4 GHz industrial, scientific and medical radio band. However, it is necessary to have an extra interface that implements Wi-Fi Direct or Bluetooth.  
	
		\end{itemize}
	\begin{figure}[t!]
		\includegraphics[width=\linewidth]{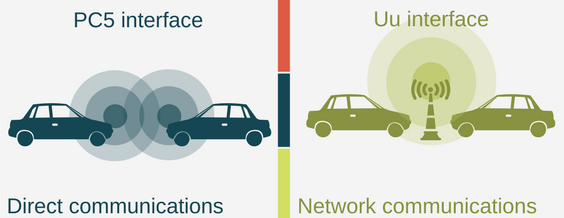}
		\caption{Communication links in LTE-V2X.}
		
	\end{figure}
\begin{figure}[t!]
\includegraphics[width=\linewidth]{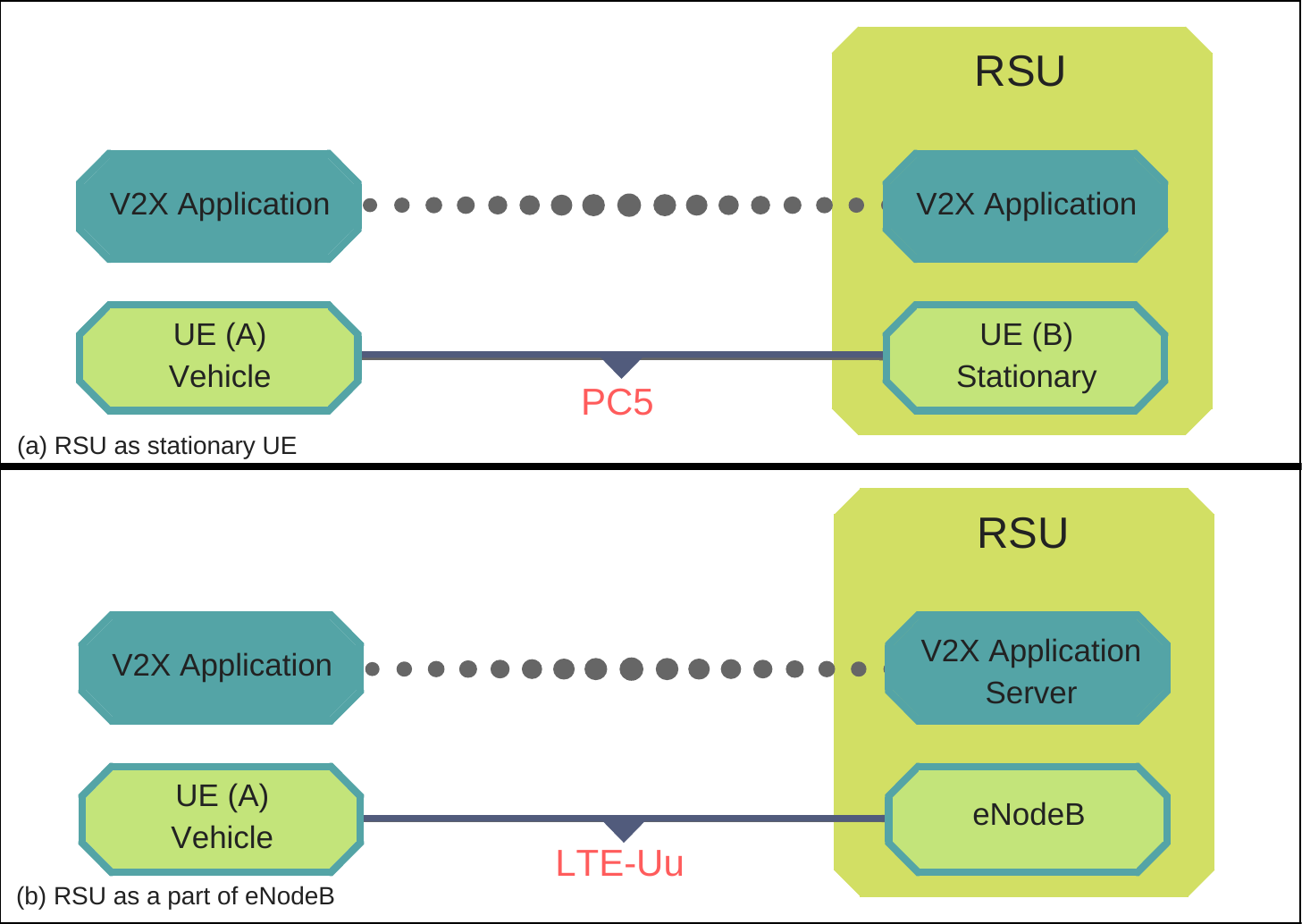}
\caption{RSU implementation in V2X.}

\end{figure}

	In LTE-V2X, RSU can be implemented in two ways (\cite{general4}). \textit{First}, it can be executed as stationary UE; then it receives V2X messages via the sidelink as shown in Figure 4 (a). In this case, the V2X application can communicate with other V2X application. \textit{Second}, it can be implemented in eNB, where it receives V2X messages via LTE radio interface as shown in Figure 4 (b). In this case, V2X application in UE communicates with the V2X application server in eNB.

\subsubsection{Security Measures} 
There are two types of LTE security mechanisms based on the type of communications as follows.
\begin{itemize}
	
	\item \textit{Cellular-based communications:} There is a mutual authentication between UE and the core network. This is achieved by using the evolved packet core authentication and key agreement procedure and generating ciphering key and an integrity key. Moreover, UEs apply different keys for different communication sessions where session keys are produced using ciphering key and integrity key (\cite{t3}).
		When an UE connects to the LTE network over LTE radio interface, the mobility management entity is responsible for executing a mutual authentication with the UE.			
	\textit{First}, the UEs send authentication requests to the mobility management entity. \textit{Second}, the mobility management entity checks the request validity and forwards it to home subscriber server, which is combined with the authentication centre, to manage and check user authenticity.
	\textit{Third}, the authentication centre generates and sends the authentication vector for specific UE to the mobility management entity (\cite{ref18}). However, the token that is sent by UE is not encrypted and not integrity protected. After mutual authenticated, the key is exchanged between the mobility management entity and UEs, then the UEs will be able to access the core network.
	UE uses a pre-shared symmetric keying and integrity algorithm for signalling packets that is sent to the network. However, for user plane data, only ciphering algorithm is applied between UE and eNB (\cite{t2}).
	
	\item \textit{D2D communications:} 
	Before starting D2D communication, UEs have to finish authentication procedure with the core network (\cite{ref17}). 
	Because the core network is responsible for managing the security parameters, the availability of accessing the network will directly affect the security level of D2D communications.
	The connectivity of UEs is based on three scenarios (\cite{ref19}). \textit{In-network coverage} when both UEs are located in the network coverage, they can communicate securely by the assist of the core network;
	\textit{partial coverage} when only one UE is in the network coverage, the communication is also managed by the core network, thus, they can establish a secure communication.
	Also, it includes when both are located at the edge of eNB coverage. They can forward their information to other covered UEs using D2D link to communicate with the eNB. In this case, the core network fully manages D2D communications as same as the "In-network coverage" scenario; 		
	and \textit{out-network coverage}
	when the UEs cannot communicate with the core network. 
	In this case, each UE saves the authentication vector which is released by the authentication centre for securing D2D communications (\cite{ref17}). The authentication parameters are valid for specific time and they might be revoked at any time, however, the security risk is increased. 
	
	To overcome this challenge, each group has a group ID, which is corresponding to the ProSe group ID (layer 2), Algorithm ID and proSe group key. Also, each UE within the group has a particular member ID (\cite{t3}). Each proSe group key is provided with an expiry time. When UE is located out-of-network coverage, the UE may work for a longer time without extra provisioning when proSe group keys valid at that time.
	The data in D2D communication is encrypted by proSe encryption key which is derived from proSe group key (\cite{t3}).
	In this case, the confidentiality requirement is achieved in data transmission. However, there is no integrity protection on the user data because the integrity key is shared by all group members.

\end{itemize}

	\subsection{Threats Analysis for IEEE 802.11p and LTE-V2X}
Similar to most communication networks, V2X security requirements can be divided into five points: availability, data integrity, confidentiality, authenticity and non-repudiation (\cite{general2}). However, in the V2X environment, attacks which targeting information availability are the most dangerous because they cause a serious effect on safety-critical situations. The following presents the security requirements for V2X network and threats against each one. Also, we present security analysis for IEEE802.11p and LTE-V2X.

\subsubsection{Threats on Availability}
They prevent the authorized users from access the information such as: 
\begin{itemize}
	\item \textit{Blackhole and Greyhole attacks:} the compromised node stops relaying packets to the neighboring nodes. Thus, it blocks up the spreading of information over the network. The attacker drops all packets that received in blackhole attack, while drops some packets in the greyhole attack.
	In IEEE 802.11p, each node must be authenticated to be a part of packets' route. In this case, the authentication process can deny the external attacker from initiate blackhole/greyhole attacks. However, the standard fails to protect the network from internal attackers. The broadcast of warning packets could reduce the effect of attack because of the diffuse multiple copies over the network.
	
	On the other hand, LTE-V2X is capable of eliminating external attackers by applying mutual authentication between UEs and the core network. In D2D communications when two UEs are located out of the network coverage, there is a possibility that UEs communicate with another UEs with revoked credentials because they are not provisioned by the core network.
	In addition, the internal attacker is possible in case of partial D2D coverage when UE that is located at the edge of eNB coverage uses other UEs to relay the packets to eNB. In this case, the relaying node could be a compromised node which drops the received packets and blocks the communication with eNB.
	
	\item \textit{Flooding attack}: 
	The attacker sends a huge volume of packets to make victim node unavailable.
	The IEEE 802.11p MAC is vulnerable to flooding attack. To initiate the attack, the attacker may exploit the binary exponential backoff scheme. Among the competing nodes, the winning node captures the channel by sending data constantly. Thus,  it causes delay in transmitting data by forcing loaded neighbors to backoff for a long time (\cite{ref20}).
	Another weakness in IEEE 802.11p MAC is network allocation vector field (\cite{ref20}). 
	When there is a communication between two nodes, the nearby nodes update their network allocation vector based on the communication duration. During that period, all neighboring nodes stop transmission and only overhear for the channel. On the other hand, the attacker may transmit bogus message to cause errors in the transmitted packets. As a result of MAC authentication, the previous attacks could only be initiated by internal attackers.
	
	In LTE-V2X, there may be two potential techniques to initiate flooding attack  against a specific UE (\cite{ref21}). 
	First, the malicious node can use the resource scheduling information to transmit an uplink control signal when another node uses the channel to transmit its information. Thus, it causes a conflict at the eNB. Second, the UE is permitted to stay in active mode, but turn off its radio transceiver to save its power resource. During that mode, the UE is still allowed to transmit packets in urgent situations. However, the attackers can inject packets during that period to cause flooding attack.
	
	Moreover, each UE transmits periodically buffer status reports to eNB which are used for packet scheduling and load balancing (\cite{ref21}). Flooding attack could happen when the attacker impersonates other UE and sends fake reports that indicate larger data volume than in the real UE. As a result, eNB may stop accepting new requests for joining the cell because it thinks that the cell is fully loaded (\cite{ref21}).
	
	  \item \textit{Jamming attack}: the attacker broadcasts signals to corrupt the data or jam the channel. 
	Both IEEE802.11p and LTE-V2X physical layer is based on the orthogonal frequency-division multiplexing technology. In fact, a jammer requires to recognize the presence of packets to launch jamming attack. However, non of current solutions can prevent jamming signals. Using directional antennas could minimize the effect of this attack and allow for vehicles to avoid the jamming area (\cite{ref15}). 
	
		\item \textit{Coalition and platooning attacks}: a group of compromised nodes collaborate to initiate malicious activities such as blocking information or isolate legitimate vehicles. 
	For instance, when several internal attackers collaborate and initiate blackhole attack, it could affect the information delivery even with broadcast nature that it is supported by the IEEE802.11p. Also, these attacks are not only limited to blackhole but include any malicious behavior. In LTE-V2X, several compromised node could prevent the traffic from the node at the edge of eNB coverage. 
\end{itemize}

\subsubsection{Threats on Integrity}
Ensuring the data integrity includes the assurance of the accuracy and consistency of data that spreading over the network. 
The following attacks address data integrity (\cite{general8}):
\begin{itemize}
	\item \textit{Alter or inject false messages attack}: the compromised node spreads bogus messages in the network, by generating a new message or modifying the received one. In both cases, it misleads the vehicles by giving them a wrong information and putting them in a danger situation.
	
	In IEEE802.11p, it is possible that internal attacker may try to broadcast false safety messages through the network. Broadcast messages are intended for all surrounding nodes, but they need to be signed in order to hinder external attackers from generating bogus messages. Because internal attacker is an authenticated node, it could use its digital certificate to sign any number of false messages (\cite{ref15}). However, the standard requires an additional scheme to be able to detect the attackers and then includes them in certificate revocation lists. In addition, these messages are protected from alternation because they are digitally signed by the sender's certificate. 
	
	In LTE-V2X, external attacker cannot inject/alter any packet because the user data in LTE-Advanced is encrypted with ciphering key which is generated after a mutual authentication. 
	On the other hand, the internal attackers can inject false information in the network. Also, they can alter the received messages because in all communication scenarios the integrity algorithm is only applied on signalling packets.
	
		\item \textit{Replay attack}: the compromised node captures the packets and replays them in a different time to look like that they were sent by the original sender.
	The vehicles which are operating in IEEE802.11p can defend against replay attacks where each node has a cache of recently received messages. Any new message is compared with the ones in the cache, and messages older than a predefined time are rejected (\cite{ref20}). 
	To prevent replay attack, LTE-V2X uses a timestamp and a nonce in each message. Also, the message has a short lifetime in the network, which leads to reject any repeated messages (\cite{ref21}).
	
		\item \textit{GPS spoofing attack}: the compromised node sends messages with fake location or after a period of time. 
	In general, GPS is responsible for delivering both location and time to the surrounding nodes. However, GPS antennas are vulnerable to damage from storms and lightning. In addition, the antennas are susceptibility to jamming or spoofing attacks. 
	In both protocols, the attackers can send fake location by generating strong signal from a GPS satellite simulator. Moreover, a successful replay attack could happen when UE uses GPS clock for message timestamp because it is easy for the attacker to spoof GPS clock (\cite{reference5}). Thus, the repeated message, which is sent after a while, could be accepted as a new message (\cite{ref15}). In this case, the protocols require to propose plausibility checks to detect fake location and time.
		\end{itemize}
	
	\subsubsection{Threats on Confidentiality}
	The confidentiality requirement allows for information to be known only by the intended receiver. This is usually achieved by encrypting the message with the public key of the receiver where it only can be decrypted by the private key of the intended receiver.
	Some threats violate this requirement (\cite{general1}), such as:
	\begin{itemize}
		\item \textit{Eavesdropping attack}:  it aims to capture packet's information and acquire sensitive and confidential information.
		Broadcast messages generally contain traffic safety information which are considered non-confidential information (\cite{ref20}).
		On the other hand,  location based services and other types of transaction application information are encrypted in IEEE802.11p. As long as the internal attacker can collect information without a permission from other users, it considers a challenge.
		
		
		In LTE-V2X, the UEs use the pre-shared secret keys which are issued by the authentication centre to establish a secure communication with the core network. Thus, the external attacker cannot collect information because the communication is encrypted.
		While in D2D communication, the UE use a pre-shared secret key for each application. However, this may cause additional computation costs on UEs (\cite{ref17}). In both communications, internal attacker can gather network information.
		To reduce the latency and computation cost, the message is not encrypted. As a result, the external attacker can collect confidential information. 
		
			\item \textit{Location tracking}: sharing the location with neighboring nodes is very important for various vehicular applications. As a result, attackers can collect and use these information for tracking users. 
		In IEEE802.11p, every time an on-board unit broadcasts message to warn neighboring vehicles regarding a safety update, it digitally signs the message with its own certificate.
		Therefore, the receiving nodes can identify the sending on-board unit and its current position. Unfortunately, wireless access in vehicular environment cannot support anonymous broadcast messages (\cite{ref15}). 
		In LTE-V2X, the core network assigns different temporary identifiers, which are continuously changed, to UEs. The temporary identifier is sent to UEs in a plain text. Thus, the passive attacker is able to identify the location of the user at that time (\cite{ref21}). However, location tracking could be achieved only if the attacker can perform the mapping between various identifiers and the UE. As a consequence, both internal and external attackers can initiate that attack. 
		
			\end{itemize}
	\subsubsection{Threats on Authenticity}
Ensuring the authenticity includes the process of giving nodes access to the network based on their identity using certificates and digital signatures. It works like a wall that protects the network from external attacks. Some threats violate this requirement (\cite{general10}) such as:	
\begin{itemize}
\item \textit{Certificate replication attack}: the compromised node uses replicated certificates to conceal itself by deleting the certificates that were added to the blacklist.

\item \textit{Sybil attack}: a single compromised node pretends many fake identities.
	\item \textit{Masquerading attack or impersonation attack}:The attacker exploits a legitimate identity to obtain access to the network and confidential information.

\end{itemize}

	Using fake identity in IEEE802.11p by the internal or external attacker could be prevented using certificate revocation lists. Each on-board units and RSUs can be identified by their certificate and in case that the normal node turns to behave maliciously, the identity will be added to the certificate revocation lists.
In LTE-V2X, the home subscriber server is able to avoid the messages that come from unauthorized UEs by applying the mutual authentication. However, the attacker could send a repeated message with a new timestamp. Because the replay attack can be detected in LTE-V2X, the home subscriber server can detect the replication and drop the message. As a result, the user impersonation or sybil attack will also be prevented (\cite{ref22}).
However, UEs are not protected when they are located out of the network coverage because they cannot ensure that they have the updated certificate revocation list. 

\subsubsection{Threats on Non-repudiation} 
Non-repudiation is responsible for identifying the real node's ID which performs a specific behavior. 
It is a method to ensure message transmission between entities via digital signature and/or encryption (\cite{general2}).
In IEEE802.11p, the periodic message and warning messages are signed with the sender certificate. Also, the location based services and any sensitive information are encrypted. 
Thus, the node identity can be identified when it launches a malicious behavior.

In addition, the message exchange between UEs and eNB is encrypted with cipher key. In addition to the encryption, LTE-V2X applied integrity protection on the signalling packets.
On the other hand, it is severe to achieve non-repudiation requirements in cooperative D2D communications because of trust concept. Indeed, if one node trusts another node, it communicates with it even if it used fake identity. Thus, that violates some features of non-repudiation (\cite{ref24}). 

\subsection{Summary} 
In this section, we provide a summary for the supported security services in each communication protocol. Table 2 summarizes the comparison between IEEE802.11p and LTE-V2X regarding to security services. We concluded that D2D link in LTE-V2X is the most exposed protocol to external and internal attacks. The reason behind is that D2D link is not managed by the core network when both nodes are located out of the network coverage. Therefore, designing a security model for D2D link is recommended to achieve high security level. In contrast, IEEE802.11p has more security services which achieve more secured ad-hoc link between vehicles.

\section{Security Solutions for V2X communications}
Vehicular communications encounter many challenges, therefore, ensuring a secure communication is considered a complex function.
Because of the shortage of the security services in both enabling technologies of V2X communications, additional security model is needed to protect vehicular network.
This section presents various security solutions that were proposed for vehicular networks. We divided them based on the security method into three categories: cryptography-based, behavior-based and identity-based. The outline of the proposed classification scheme is shown in Figure 5.
	\begin{table}[b!]
	
	\caption{Comparison between IEEE802.11p and LTE-V2X regarding to security services}
	\label{multiprogram}
	\scalebox{0.7}{\begin{tabular}{|c|c|c|c|c|c|c|c|c|c|c|c|}
			
			\hline
			&  &  \multicolumn{2}{c|}{\textbf{IEEE802.11p}} & \multicolumn{4}{c|}{\textbf{LTE-V2X}} \\ 
			\cline{3-8}
			
			\makecell{\textbf{Security} \\ \textbf{Requirements} } & \textbf{Threats} &  &  & \multicolumn{2}{c|}{Cellular-based} & \multicolumn{2}{c|}{D2D-based} \\
			\cline{5-8}
			\multicolumn{1}{|c|}{} && External & Internal & External & Internal & External &Internal  \\
			\cline{1-8}	
			\multirow{3}{*}{Availability}  &  
			\makecell{Blackhole \\ \ Greyhole attacks} & \cmark   &\xmark&\cmark&\xmark&\xmark&\xmark\\ 
			\cline{2-8}
			&   Flooding attack & \cmark &\xmark &\cmark&\xmark&\xmark&\xmark\\
			\cline{2-8}
			&   Jamming attack & \xmark&\xmark&\xmark&\xmark &\xmark&\xmark\\
			\cline{2-8}
			&   Coalition attack & \cmark &\xmark&\cmark&\xmark &\cmark&\xmark\\
			\cline{2-6}
			\hline
			\multirow{4}{*}{Integrity}  &  
			\makecell {Alter messages attack} &\cmark  &\cmark &\cmark&\xmark&\cmark&\xmark\\ 
			\cline{2-8}
			&\makecell {Inject false \\ messages attack} &\cmark  &\xmark &\cmark&\xmark&\cmark&\xmark\\ 
			\cline{2-8}
			&   Replay attack & \cmark&\cmark&\cmark&\cmark&\cmark&\cmark\\
			\cline{2-8}
			&   \makecell {GPS spoofing attack} & \xmark &\xmark&\xmark&\xmark&\xmark&\xmark\\
			\hline
			\multirow{2}{*}{Confidentiality}  &  
			Eavesdropping attack&\cmark  &\xmark&\cmark&\xmark&\xmark&\xmark\\ 
			\cline{2-8}
			&   Location tracking &\xmark  &\xmark &\xmark&\xmark&\xmark&\xmark\\
			\cline{2-8}
			\cline{2-8}
			\hline
			\multirow{5}{*}{\makecell{Authenticity}}  
			&   \makecell{Certificate \\ replication attack}&\cmark  &\cmark&\cmark&\cmark&\xmark&\xmark\\  
			
			\cline{2-8}
			&Sybil attack &\cmark &\cmark&\cmark&\cmark&\xmark&\xmark\\ 
			\cline{2-8}
			&   \makecell{Masquerading \\ /impersonation attack}  & \cmark&\cmark&\cmark&\cmark&\xmark&\xmark\\
			%

			\hline
			Non-repudiation	& Any  & -&\cmark&-&\cmark&-&\xmark\\
			
			
			
			\hline
	\end{tabular}}
	
\end{table}

\subsection{Cryptography-based Solutions} Cryptography is responsible for achieving a secure communication between sender and receiver by developing protocols that prevent unauthorized users from accessing the network (\cite{reference7}). As a result, these solutions concentrate on external attacks that are launched by unauthorized users.
\begin{figure*}[t!]
	\includegraphics[width=\linewidth]{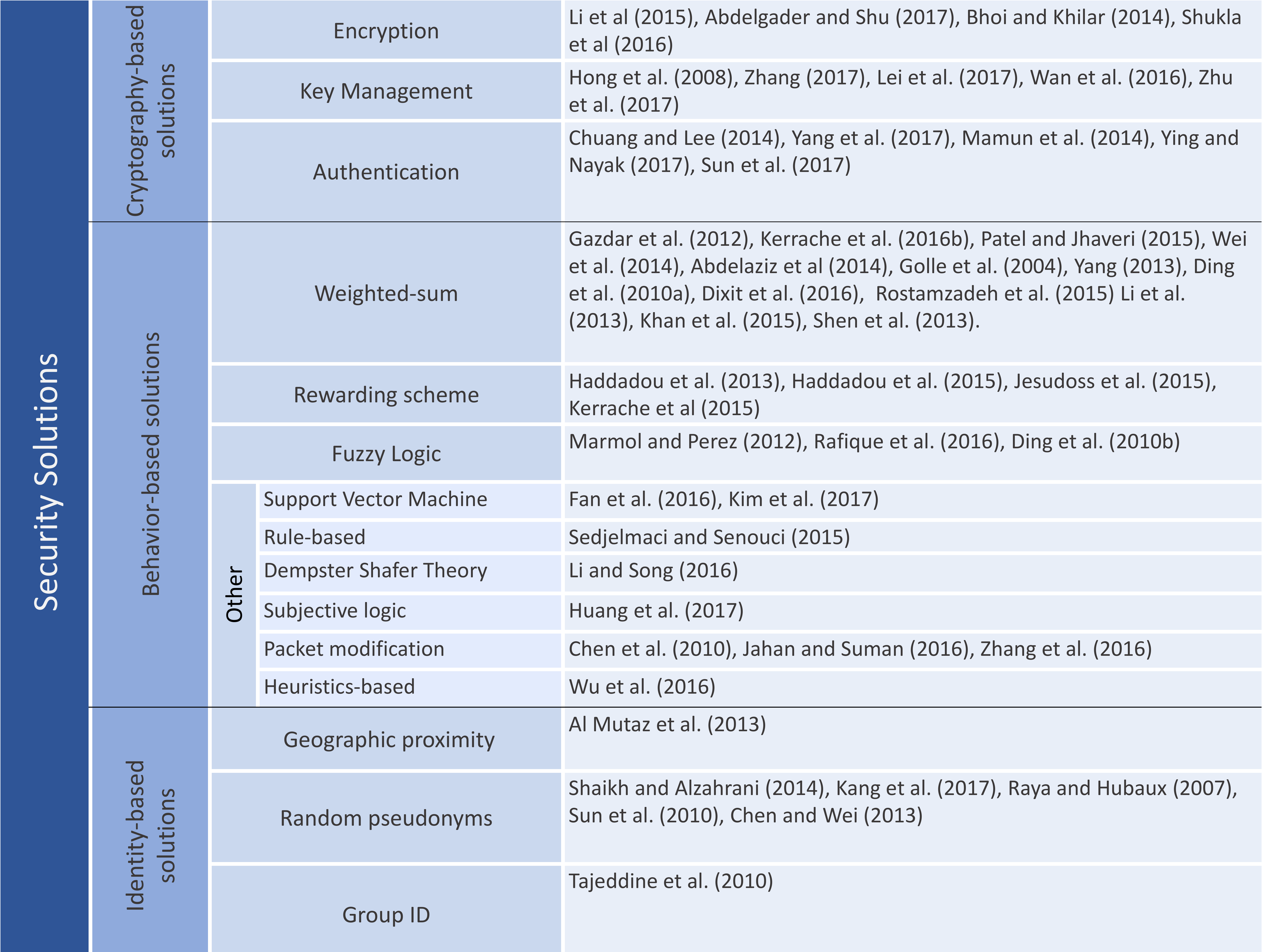}
	\caption{Overview of the surveyed research works - Classified according to the security methods.}
	
\end{figure*}
\textit{Encryption} is the main technique in cryptography-based solutions where it uses various algorithms to transform the data into another form that is only readable by intended users (\cite{general7}).
For example, \cite{vanet32} proposed a lightweight secure navigation system for vehicular ad-hoc network. Each vehicle applies encryption and the digital signature for supporting a secure communication between the vehicle and RSU. The system addressed the replay attack and man-in-the-middle attack. In addition, the proposed model in (\cite{ref3}) applied advanced encryption standard to achieve the user privacy. The key distribution problem of advanced encryption standard is addressed by taking advantage of the randomness of the channel in vehicular networks to share the secret key.
In addition, a Secure and Intelligent Routing protocol (\cite{vanet5}) used a double encryption for the data packets and applied authentication scheme to measure node's trust. However, it caused an increase in the processing time and raised the network overhead.  

\cite{ref5} proposed a model that obtains the security in the vehicular network using multiple operating channels where the encrypted data is sent on multiple channels. The receiver considers the mean value of the received data. For example, if the attacker hacks some channels and alters the data. Then,  the partial incorrect data is received which is improved partially by taking the mean of the received data. Also, in case of jamming of particular channels, the receiver still can receive the data using other channels. However, this method is only considered when the security is a major concern rather than resources because it achieves a high bandwidth waste.

Key management process manages the cryptographic keys in a crypto-system which includes the generation, exchange and storage of the keys. For instance, \cite{vanet39} proposed a situation aware trust that is based on vehicle situations. For building the situation aware trust, it first defines the attribute (static or dynamic) such as time, company and location. The group which belong to specific attribute share the same key. Then, the packet is encrypted using that key where it only can be read by the group members. For example, if there is a taxi from company A and want to send a message to other drivers. The message has four attributes as follows (Company, Taxi, Time, Location). As a result, only vehicles with these attribute can read the message.
Moreover, \cite{ref2} proposed a novel method based on one-time identity-based authenticated asymmetric group key agreement to create cryptographic mix-zones which resist malicious eavesdroppers. The safety messages are encrypted using a group secret key to improve vehicle privacy. Thus, any external entity cannot track the safety messages in the cryptographic mix-zones. 
In addition, \cite{ref11} proposed a novel key management scheme for key exchange among security managers in heterogeneous networks. They used blockchain concept for allowing key exchange securely within the security manager network. Flexible transaction collection period was proposed to reduce the key exchange time in blockchain scheme.

A key generation model in (\cite{vanet34}) used received signal strength to guarantee the randomness in key generation. The node senses the received signal and transforms the received signal strength values to binary value by applying upper threshold ($Th_{up}$) and lower threshold ($Th_{low}$). For key generation, the node sets one if the received signal strength value greater than $Th_{up}$ and zero if the received signal strength value less than $Th_{low}$. Any value in between is ignored. Also, \cite{ref8} proposed a scheme which grants the vehicles the ability to generate a shared secret key from received signal strength indicator values with low probability of getting the same key by neighboring vehicles.

\textit{Authentication} schemes were proposed using various techniques to detect unauthorized nodes and prevent them from launching malicious attacks. For instance, \cite{vanet22} introduced a scheme for authenticating the vehicles by considering them trusted nodes if they are successfully authenticated. On the other hand, \cite{vx1} proposed two lightweight anonymous authentication schemes for the V2X network. One scheme was applicable for V2V communication, while the other was suitable for V2I communications. Both schemes were considered the limitations in V2X such as resource constraints of on-board unit and latency limit.

\cite{ref7} presented a reliable and standard chosen plaintext secure group signature solution for vehicular network applications. It allows any fixed entities such as RSUs to link messages, and recognize if they are generated by one or group of vehicles, without breaching their privacy. Thus, it prevents malicious complaints against normal nodes. 
In addition, \cite{ref10} proposed an anonymous and lightweight authentication based on smart card protocol. It consists of two main phases which are user authentication and data authentication. It protects the network against various attacks such as offline password guessing attack, impersonation attack and many others. Also, the anonymity is achieved by using dynamic identities.

The work of \cite{ref14} applied a model which achieved the privacy requirements. It is composed of two schemes which are identity-based signature and pseudonym scheme. Also, it provides the authentication in inter-vehicle communications.

\subsection{Behavior-based/Trust-based Solutions} They were proposed as complementary solutions to cryptography, where traditional cryptography-based solutions are not able to detect internal attackers because they are authenticated users (\cite{general8}). 
The trust evaluation is typically conducted by monitoring nodes, which monitor and collect other nodes' behavior information. The most common methods that are used for trust management in vehicular networks as follows:

\begin{itemize}
	
	\item \textbf{The weighted-sum method} is the most common methodology for trust management, where trust evaluation is computed by assigning different weights for each trust component. When the node behaves maliciously; the total trust value decreases until reach to zero (\cite{general6}). Total trust is computed by:
	\begin{equation}
	T_{total}  = \sum_{i=1}^U w_i \times T_x
	\end{equation}
	
	where $w_i$ is a weight value for $T_x$, $T_x$ is a trust value for trust level $x$ such as direct and indirect. Indeed, direct trust measures trust level of one-hop neighbors using direct monitoring, while indirect trust measures trust level of two-hops neighbors using the recommendations from other nodes. $U$ is the number of trust levels that will be considered.
	
	\cite{vanet1} proposed a dynamic and distributed trust model for vehicular adhoc network that used the monitoring nodes for observing their neighboring nodes and sending an alert about any malicious activity such as packet dropping or packet modification.
	On the other hand, \cite{vanet17} proposed trust model that provided two trust metrics: vehicle-based and RSU-based. Each node is able to monitor its neighbors and measure local trust value. Then, RSUs are responsible for managing a global and historical trust information about all nodes which locate in the same road segment. At the same time, the model could work without the existence of RSU.
	\cite{vanet6} applied ant colony optimization algorithm for choosing the shortest trusted path by isolating non-cooperative nodes. However, the node is compelled to transmit the packets to the next hop even if all neighbors are malicious. Unlike (\cite{vanet6}), \cite{vanet2} proposed a trust model for detecting non-cooperative nodes on V2V communications only. Moreover, \cite{vanet18} proposed intrusion detection system to enhance the message relay mechanism by evaluating the trustiness of received messages.
	
	Event-based models collect data and monitor different events going on in the environment to build the reputation. In the vehicular networks, some models were proposed to monitor traffic events and evaluate the trustworthiness of these events. For example, event-based reputation model was proposed in (\cite{vanet26}) for checking data consistency. Indeed, the sensors provide redundant information which allowed for each node to process the data and remove malicious information. If inconsistencies occur, the security model is activated to detect the malicious node. Furthermore, similarity mining technique was suggested in (\cite{vanet30}) for recognizing similarity among vehicles and messages. For instance, at a similar location and similar time, messages about the same event that are generated by the same vehicle usually have similar trust values. In addition, \cite{vanet21} proposed event-based reputation model to filter fake warning messages and increase the accuracy of the network. It measures the reputation of the event based on its roles (event reporter, event observer or event participant) to check if the event triggered is real or fake alarm. 
	Furthermore, trust value can be related to a specific location. 
	For example, when trustworthy interactions are established between vehicles in a specific road segment, then, the corresponding trust value for that location and time is increased.
	\cite{vanet15} proposed a model for selecting the trusted location using Ad Hoc On Demand Distance Vector routing protocol where RSUs were responsible for controlling packets routing. However, as the number of malicious nodes increased, the vehicles may follow the wrong path. In addition, the proposed framework in (\cite{vanet14}) composed of two modules: the first one implements three security checks to measure the message's trustworthiness. First, it examines that the message is generated from a trusted location and followed a trusted route. Second, it inspects that message route does not contain any malicious nodes. Third, it checks that it has a valid content. Indeed, it computes a trust value for each road section and for each neighborhood. Once a message is evaluated and considered trusted, then it looks up for a trusted route to forward the message.
	
	  The most recent malicious nodes behave intelligently and conceal themselves from detection by alternating between normal and malicious behaviors. To protect the network from smart attacks, \cite{vanet13} offered a Reputation-based Global Trust Establishment to address a smart blackhole attack by considering the past behavior of nodes.
	  
	  The monitoring role could be concentrated on the message content. The monitor nodes measure trust value based on the validity of the received messages. For instance, \cite{vanet9} proposed an algorithm for vehicular ad-hoc networks which provides distributed message authentication. It gave the monitoring roles to specific nodes called verifier nodes. These nodes were responsible for verifying the message and transmitting the decision through the network. The model considered three parameters for choosing appropriate verifiers which are load, distance and trust value.
	  In addition, \cite{vanet19} proposed a distributed authentication scheme where verifier vehicles were responsible for checking the validity of the message while non-verifier vehicles depend on verification results. The choice of verifier nodes is made using three methods: N-nearest method, most-even distributed method and hybrid method.

	  \item \textbf{The rewarding-based method} uses credit to reward cooperative nodes. For instance, the node rewards its neighboring node while it behaves normally and cooperates with other nodes. Thus, a node with high trust value considered a reliable node. The rewarding method is used as a security solution to encourage non-cooperative nodes to participate in packet forwarding process. For example on this, \cite{vanet3} and \cite{vanet12} proposed a rewarding scheme for detecting blackhole and greyhole attacks.
	  Moreover, \cite{vanet41} proposed a Payment Punishment Scheme where it is applied during election and routing processes. They used vickrey, clarke and groves model and designed the payments in a way that the collaborative node will gain rewards and be able to participate in election and routing processes. In addition, the model assigns monitoring nodes which are responsible for monitoring the behavior of the relay nodes.
	  The work in (\cite{vanet10}) represents trust-based routing protocol which is composed of two main phases. The first phase includes the trust measurement, which is done periodically and in distributed manner. When the node receive a data packet, it applies trust model on the received packet. Thus, it is able to evaluate the sender behavior and compute its trust value. Then, based on the previous step, the second phase involves sending the data through the most trusted route.
	  
	  	\item \textbf{The fuzzy logic method} incorporates a series of IF-THEN rules to solve a control problem rather than attempt to model a system mathematically. The main steps of the fuzzy logic model are as follows (\cite{fuzzy1}). First, the fuzzy sets and criteria are defined; next, the input variable values are initialized; then, the fuzzy engine applies the fuzzy rules to determine the output data and evaluate the results.	
	  For instance, \cite{vanet29} proposed a security model that worked on detecting selfish nodes that transmit false or bogus messages. The model defined a fuzzy set to classify each node with three different trust levels. Based on the source node trustworthiness level, the receiver can decide whether it has to receive, forward or drop it.
	  Also, fuzzy logic models were proposed in (\cite{vanet7}) to detect packet dropping attack. Moreover, \cite{vanet27} proposed a fuzzy reputation based model to prevent the spreading of false messages.
	  
	  	\item \textbf{Others methods}
	  Some existing models utilized different methods to detect internal attacks. For example, \cite{vanet31} proposed a detection system using support vector machine learning algorithm to stop replay attack. It applied a training phase to fetch attack's characteristics. In addition, \cite{ref1} proposed collaborative security attack detection mechanism in a software-defined vehicular cloud architecture. It consists of two phases: \textit{information aggregation phase} where each vehicle analyses the received information and transmits the result periodically to the controller for training the support vector machine, 
	  and \textit{Multi-class support vector machine training phase}. Then, the classification using support vector machine is started where each vehicle applied the classifier to detect malicious nodes. However, this method is not energy-efficient because it requires training phase to be able to detect the attack. 
	  \cite{vanet11} proposed a framework as intrusion detection system that concentrated on behavioral attacks. It addressed various attacks in vehicular adhoc networks such as selective forwarding, blackhole, wormhole and sybil attacks. It uses special agents which are responsible for monitoring the nodes' behavior and triggering an alarm when the misbehavior is detected. It is composed of two detections systems and decision system. First, local intrusion detection system which operates on each node to monitor its neighbors and the cluster head.  Second, global intrusion detection system which runs at the cluster head level to monitor its cluster members. Finally, global decision system runs at RSU level to calculate trust level for each node by aggregating the reputation of each node and broadcast the blacklist through the network.
	  
	  In addition, \cite{vanet37} proposed a model using dempster-shafer theory of evidence to combine multiple evidence even if some of them might not be accurate. It was proposed to evaluate two types of trust: data trust and node trust. 
	  In particular, data trust is used to evaluate the received data and indicate if it is acceptable based on a calculated trust value. On the other hand, node trust assesses the node based on its behavior with neighboring nodes and indicates whether or not it is trusted. 
	  
	  A distributed reputation management system was proposed in (\cite{ref13}) for securing vehicular edge computing. Vehicular edge computing servers are used to implement local reputation management tasks for vehicles. In addition, they applied multi-weighted subjective logic for computing the reputation values.
	  On the other hand, some existing solutions implemented adjustments on routing protocol to increase the security level.
	  For instance, in the model that was proposed in (\cite{vanet20}), each node appends a list of trust opinions with the cluster data. Also, it applies confidence level for its opinion to determine the assurance level about the computed trust value. Moreover, \cite{vanet4} proposed a change in routing protocol to detect non-cooperative nodes using a double acknowledge technique. In another work, \cite{vanet8} proposed an amendment to adhoc on-demand distance vector routing protocol to identify packet dropping behavior by adding new fields to the control packets. The main drawback of this method is the risen network overhead as a result of the increase in packet size.
	  
	  Moreover, \cite{reference6} proposed MobiFish which is a lightweight anti-phishing application. They develop enhanced version of heuristics-based method by addressing the high relying on web page source code. They implemented it on a Google Nexus 4 smartphone running the Android 4.2 OS.
	  
	\end{itemize}
\subsection{Identity-based Solutions} They address attacks that exploit users' identity for malicious activities such as sybil attack and breach user's privacy. 
Most security solutions used the vehicles€™ identity to identify them and revoke malicious node. As a result, user's privacy has been revealed and misused by an attacker. To resolve this issue, the security model should operate on an identity anonymous environment. For example, \cite{vanet33} proposed the first trust management scheme which takes into account the anonymous identity. It protects vehicular ad-hoc networks from spreading messages with fake location and time. Each node calculates its confidence value regarding the received messages about a specific event. The confidence value is based on four parameters: location closeness, time closeness, location verification and timestamp verification. Then, it calculates the trust value for each message that reporting the same event and makes the decision regarding that message based on the trust value.

 Moreover, \cite{vanet36} proposed a privacy-preserving trust framework that allowed for collecting information about vehicles' behavior while preserving their privacy by applying a group ID rather than real identity. Also, \cite{vanet35} proposed a beacon-based trust model that ensured the vehicular ad-hoc networks safety while preserving the drivers' privacy. All transmitted messages were protected by cryptography and the pseudo identity schemes.

One common approach to preserving the privacy of vehicles' location is the use of pseudonyms. Indeed, the vehicle broadcasts its information with pseudonyms that frequently changes (\cite{vanet52}). As an example of this, \cite{CV4} proposed a defend system for two cases of eavesdropping. The first case, an adversary tracks a target vehicle by a specific pseudonym. In this case, the security solution used a random Virtual Machine identifiers which make the mapping relationship fail. Unfortunately, the first solution was vulnerable to identity mapping attack. Because of that, they proposed Pseudonyms Changing Synchronization Scheme to improve detection accuracy. 
In addition, \cite{ref9} presented the vehicular ad-hoc network security system which based on three main techniques. 
First, a pseudonym-based technique which was used to assign pseudonym/private key pairs to the vehicles which are traveling in the home domain or other domains. Second, Threshold signature which was applied to send the secret information for recovering a malicious vehicle's identity. Meanwhile, it prevents compromised nodes from having a full authority to revoke a normal node. Third, Threshold authentication based defence scheme which provides a mechanism to distinguish between malfunctioning and malicious behavior.

Geographic proximity technique is used to detect sybil attack when malicious node uses multiple identities. \cite{vanet45} applied this method to identify sybil identities where the geographic proximity of the compromised node and all its sybil identities last for a long time and repeated.

\subsection{ Summary}
In this section, we provide a summary to get an overview of the proposed security methods. Table 3 summarizes characteristics of the main security models were proposed for vehicular networks. The parameters of this summary are presented as follows:
\begin{table}[b!]
	
	\caption{Main security solutions in vehicular networks}
	\label{multiprogram}
	\begin{threeparttable}
		\makebox[400pt]{
	\scalebox{0.60}{\begin{tabular}[2ex]{|c|c|c||c|c|c||c|c|c||c|c|c||c|c|c|c|c|}
			
			\cline{1-17}
			& \multicolumn{5}{c||}{Topology} & \multicolumn{3}{c||}{Purpose} &  \multicolumn{3}{c|}{Addressed Attacks} & \multicolumn{5}{c|}{Security Requirements}\\
			\cline{2-17}
			& \multicolumn{2}{c||}{Organization}  & \multicolumn{3}{c||}{Architecture (Use of RSU)}& 	\multirow{2}{*} {\makecell{IDS}} & \multirow{2}{*} {\makecell{Secure \\ Route}} & \multirow{2}{*} {\makecell{Secure \\ Content}} & \multirow{2}{*} {\makecell{Internal \\ attack}} & \multirow{2}{*} {\makecell{External \\ attack}} & \multirow{2}{*} {\makecell{Both}} &\multirow{2}{*} {\makecell{Av.}}&\multirow{2}{*} {\makecell{Int.}}&\multirow{2}{*} {\makecell{Conf.}}&\multirow{2}{*} {\makecell{Auth.}}&\multirow{2}{*} {\makecell{NRep.}}\\
			\cline{2-6}
			
			\multicolumn{1}{|c|}{} &Flat  & Clustered & Centralized & Distributed & Hybrid &  &  & & & & &&&&& \\
			\hline
			\cite{vanet11}& &\cmark&\cmark&&&\cmark& & & \cmark&&&\cmark&\cmark&&&  \\
			\hline
			\cite{vanet12}& \cmark&&&\cmark&&& \cmark & & \cmark&&&\cmark&&&&  \\
			\hline
			\cite{vanet14}& \cmark&&\cmark&&&& \cmark&\cmark & \cmark&& &\cmark&\cmark&&& \\
			\hline
			\cite{vanet17}& \cmark&&&&\cmark&& \cmark&\cmark & \cmark&& &\cmark&&&& \\
			\hline
			\cite{vanet22}& \cmark&&&\cmark&&& \cmark&\cmark & &\cmark& &&\cmark&&\cmark& \\
			\hline
			\cite{vanet23}& \cmark&&\cmark&&&& &\cmark & &\cmark& &&\cmark&&& \\
			\hline
			\cite{vanet40}& \cmark&&&&\cmark&& \cmark& & &&\cmark &\cmark&&&&\cmark \\
			\hline
			\cite{ref2} & \cmark&&\cmark&&&& &\cmark &  &\cmark& &&&\cmark&\cmark& \\
			\hline
			\cite{ref8}& \cmark&&&\cmark&&& &\cmark &  &\cmark&  &&&\cmark&&\\
			\hline
			\cite{ref9}& \cmark&&\cmark&&&\cmark& & &  &\cmark& &&&\cmark&\cmark&\cmark \\
			\hline
			\cite{vanet48}& \cmark&&&&\cmark&&\cmark & &  && \cmark &\cmark&&&&\\
			\hline

			\cite{ref11}& \cmark&&\cmark&&&\cmark& & &  &\cmark& &&\cmark&&& \\
			\hline
			\cite{ref13}& \cmark&&&\cmark&&\cmark& & &  \cmark&&  &\cmark&&&\cmark&\\
			\hline
			\cite{ref14} & \cmark&&\cmark&&&\cmark& & &  &\cmark& &\cmark&&\cmark&\cmark& \\
			\hline
			\multicolumn{17}{l}{%
				\begin{minipage}{16.5cm}%
					\tiny *Av: Availability, Int: Integrity, Conf: Confidentiality, Auth: Authentication and NRep: Non-repudiation
				\end{minipage}%
			}\\
			
	\end{tabular}}}

\end{threeparttable}

\end{table}
\begin{itemize}
	\item Organization: indicates whether the model applied on flat or clustered network. 
	\item Architecture: indicates whether the model used the centralized structure, distributed or hybrid.
	\item Purpose: indicate whether the aim of the model is intrusion detection system, achieve a secure route or protect message content.
\item Addressed attacks: indicates which attacks is addressed in the model.
\end{itemize}

 We concluded that the most research applied the security model on a flat network where all nodes have the same responsibility.  In addition, many security solutions focused on addressing one type of attacks: internal or external. Using central unit for security measurements was frequently used. However, centralized models are not applicable in vehicular networks where the node could be located out of the network coverage.

\section{Comparison and Discussion}
In this section, we discuss the effectiveness of security methods for protecting V2X communications. Based on our analysis, we noticed that most of proposed models were applied on vehicular ad-hoc network where the security model was implemented on vehicles (homogeneous network). Because V2V is part of V2X communications and both of them support the communications in vehicular network, they have common characteristics such as various speeds, non-stable connection and dynamic topology. As a result, studying the security solutions in V2V is required to address their challenges while designing security model for V2X communication.
In addition, few proposed security models (\cite{vanet9,vanet11,vanet20,vanet41}) considered clustering vehicular network and all of them are behavior-based solutions.

After analyzing the existing solutions, we evaluated them based on four parameters as follows:
\begin{itemize}
	\item \textit{The considered attack:} study the ability of various security methods in protecting the network against internal or external attacks.
	\item \textit{The message type:} study to which extent the security method is important for delivering various V2X messages.
	\item \textit{The latency limit:} study the effect of applying various security methods on message delivery time.
	\item \textit{The security model structure:} study the impact of security method on the model structure.
\end{itemize}

\subsection{The considered attacks}
Some existing works (\cite{ref5,vanet5,vanet22}) used traditional security scheme for protecting vehicular networks such as encryption and authentication. These methods are important in protecting the network from external attacks, and also they achieved high-security levels in a centralized network. However, they are not reliable solutions for distributed networks. 
Moreover, based on the study on V2V applications which was done in (\cite{comment1}), they listed some limitations of various cryptography-based methods. Symmetric key systems causes additional overhead and delay in key distribution phase. While, asymmetric cryptography is suitable for distributed systems where nodes is highly dynamic. However, it is slower than symmetric key systems and causes a huge latency. Furthermore, the revocation process in group signature method needs for high computation power and also the signature is too large to transmit over the air.

Based on the previous comparison between the main vehicular network communication protocols, we noticed that the most external attacks are considered by protocol security services. Thus, applying additional cryptography model will increase the overhead on the network.


Behavior-based solutions can be implemented as a supplemental solution to fill the gap of classical cryptography solutions, as proposed in (\cite{CV1,CV2}). 
They commonly are necessary against internal attackers which they own legitimate certificate. Also, they are mostly applied on distributed and semi-centralized network (\cite{general2}). 
The weighted-sum method has some challenges such as setting the weights and trust threshold, however, it is considered a common and lightweight method for vehicular networks. Because the vehicular networks require low latency in message delivery, a lightweight security method is recommended.
Furthermore, there are a few works recommended a fuzzy logic for the vehicular network because it needs for a training phase, and it is suitable for the predictable environment. On the other hand, the rewarding-based method is suitable for encouraging non-cooperative nodes. Therefore, it limited to address the selfish behavior attack.

In addition, the shortcoming of identity-based solutions is the focus on attacks that exploit user's information to track them or pretend false identity. Thus, it is ordinarily used as complementary solution to protect user information where both protocols were not supported anonymous identity. As a result, the need for designing identity-based solutions is increased.


Some researchers made efforts to address the previous limitation by applying hybrid system which combines multiple security methods to increase the security level of the network. For instance, \cite{vanet40} utilized Ant Colony Algorithm with graph model to detect internal attacks of routing control packets. It considers the external attacks by applying a digital signature. Also, it addresses internal attacks using plausibility checks such as QoS, link breakage and control messages broadcast. 
Moreover, \cite{vanet48} suggested a security solution that combined a distributed plausibility checks and digital signature technique. The model used time stamps, transmission range and vehicle€™s velocity to detect false position injection.

\subsection{The message type}
\subsubsection{The event-based messages}
The message encryption is applied to keep the message content confidential while traverse through multi-hop route. However, the encryption is not essential for event-based message because it is directed towards all nodes in the area and not targeting specific destination. Thus, the message should be delivered to all nodes in a plain text to be able to read its content and make a decision regarding the traffic in a short period (\cite{comment1}). As a consequence, the encryption could affect negatively on the network performance. On the other hand, unauthorized nodes may start behaving maliciously and mislead other nodes by sending false messages. As a result, all nodes should be authenticated to be able to generate message to protect the network from false alarm which is injected by the external attackers. In addition, generating false alarm is not limited to external attackers but it can be initiated by authorized nodes. Much research (\cite{vanet1,vanet9,vanet14,vanet29}) were proposed behavior-based solutions for securing packet content and detect internal attacker.

The risk of tracking users by mapping event-based messages is too low because the road entity will generate a message when an event is triggered. Thus, the attacker can know the current location of the road entity but he has to wait for next event to know its next location. As a result, hiding user's information is not critical.

\subsubsection{periodic messages}
As we mentioned in section 2, the beacon message contains user's information such as location, speed and direction. This information is targeting on the nodes which are belonging to the network. Thus, this message should be encrypted to prevent external attacker from tracking users.

Moreover, internal attackers may send false status information to mislead neighboring nodes. For instance, the model in (\cite{vanet33}) proposed to detect beacon messages with false location and time. Moreover, some recent works (\cite{vanet35,vanet55}) suggested beacon-based trust where the node's trust value is measured based on the validity of beacon message.

As a result of missing mechanism in vehicular network protocols which supports user privacy, applying identity-based methods is necessary for beacon messages. The attacker can follow the beacon message and track the user's location because it is sent periodically and has much information about the user. Applying random identities is one of the main solutions for user privacy, however, it is vulnerable for mapping attack. The model in (\cite{CV4}) proposed Pseudonyms Changing Synchronization Scheme to improve the user privacy.

\subsection{The latency limit}
As we know that vehicular network is latency sensitive because the information should be delivered to other nodes in short time approximately 300 ms. As a consequence, applying complicated model will increase the delivery time. 
The main communication protocols for vehicular network support encryption and authentication services. Thus, applying additional cryptography solutions increase the packet delivery time. For instance, the model in (\cite{vanet5}) proposed double encryption scheme which increase the computation overhead. 

Behavior-based solution were proposed as a lightweight solution, however, some research work (\cite{vanet2}) were proposed complex mathematical model to measure trust value.
Also, some of these solutions apply machine learning algorithms such as support vector machine in (\cite{vanet31,ref1}) where the complexity is high for training phase and testing phase. 

Identity-based solutions are lightweight where it only based on random pseudonyms (\cite{vanet33,vanet35,CV4,ref9}) or group ID (\cite{vanet36}). These methods have simple calculations which not require much time to apply them.

 \subsection{The Model structure}

The most cryptography-based solutions applied centralized structure (\cite{vanet23,ref2,ref9,ref10}). However, centralized model is not suitable for the vehicular network because the central units such as RSUs are not always available along the roadside. Because of that, (\cite{CV3,vanet19}) proposed distributed authentication schemes. In addition, \cite{CV3} proposed distributed authentication scheme where if the node would like to request a cloud service, it sends a message containing its own identity and its location to neighboring nodes; then, the neighboring nodes authenticate the message using reputation list. However, it concentrated on V2V communication.

Behavior-based solutions designed to give the nodes ability to manage security model independently from the central unit. However, some models used central units for gathering trust values from vehicles such as (\cite{vanet7,vanet11,vanet15}). The main drawback in distributed structure is deficient in global knowledge about the network. For example, if $vehicle\ A$ meet $vehicle\ B$ for the first time, $vehicle\ A$ don't have information about the security record of $vehicle\ B$.

The proposed model in (\cite{vanet17}) applied a hybrid system which manages both model structures. The distributed structure applied when the vehicles are located in the area that is not covered by a central unit. However, this assumption gives us non-stable detection accuracy.

\section{Challenges and Research Direction}
The security in the vehicular network is still considered open research area. vehicular ad-hoc networks gave researchers the opportunity to study and evaluate their works. As a part of communication progress, the future vehicular network will not be limited to vehicles and RSUs, but it will include all roadside entities. As a result, applying traditional solutions in the V2X network cannot perform as it is expected. 

Major efforts have been made; however, several open issues are still needing more consideration to achieve a high-security level for V2X communications. Some of these issues are discussed in the following section.

\subsection{Security attacks}
In trust-based solutions, the evaluation is based on monitoring neighbor's behavior. 
As a result, they are susceptible to trust attacks, such as whitewashing attack, on-off attack and good/bad mouthing attacks. In these attacks, the compromised node behaves smartly to conceal itself from being detected by alternate between normal and malicious behavior. Unfortunately, most existing works did not address them in vehicular networks except the proposed model in (\cite{vanet13}) considered intelligent compromised nodes which behave maliciously for intermittent periods. 

The wireless communication is the prospective method to facilitate the communication inside large public transportation systems such as trains, metro and buses. However, it is more susceptible to various cyber attacks than the wired communication. Therefore, implementing security model for intra-vehicle subnetwork is recommended. For instance, \cite{vanet53} proposed secure system based on host identity protocol. It is designed to protect intra-vehicular communication from common IP based attacks.

The most centralized security solutions used RSU as a fully trusted unit for updating security records. Thus, the vehicles require communicating with RSU to get updated information. However, this assumption is not applicable in vehicular networks because RSUs are not always available along the roadside. Also, RSUs, like any road entities, are vulnerable to various attacks (\cite{general15}). As a result, in case of RSU attacks, these solutions are not working efficiently and have a huge impact on network performance. For instance, \cite{vanet54} developed security protocols for the key distribution, which are able to detect the compromised RSUs and their collusion with the malicious vehicles.

\subsection{Security vs. QoS management}


The term Quality of Service (QoS) is used to represent the level of performance provided to users. In traditional networks, high levels of QoS can be obtained by resource allocation and sufficient infrastructure. However, the control over the network resources, such as bandwidth, equipment, power consumption and transmission delay, is difficult in the vehicular networks because the network lacks of consistent infrastructure and stable topology. For instance, \cite{rev3} proposed method to fuse vehicle spacing information and estimate average traffic density estimation. However, some security mechanisms can be added to the message frame to prevent attack on the spacing information. Therefore, achieving the trade-off between QoS and security level in the vehicular networks is an essential task. 
The security model should always consider the efficient use of devices' resources because the road entities in V2X communication have various capabilities and resources.  An example of this is the power consumption, all current security models did not consider the power consumption as a challenge because they were implemented on vehicles which have a long life battery. However, in V2X, some road entities such as mobile phones have a limited battery. As a result, designing a lightweight security model is recommended for V2X. 

In addition, QoS includes the useful use of the network resources such as the bandwidth. Indeed, \cite{rev2} proposed a two-stage delay-optimal scheme by integrating software defined networking and radio resource allocation into an LTE system for vehicular networks. Also, the proposed model in (\cite{ref5}) used multiple channels to send the same information which achieves the availability but wastes the bandwidth. In addition, the proposed models in  (\cite{vanet4,vanet8}) caused further overhead by increasing the packet size.

For critical applications such as safety-related applications in vehicular networks require a minimum delay for delivering warning messages to obtain a high safety level. However, the most encryption methods need a time for encryption and decryption processes. Thus, they protect the information and achieve confidentiality but they have a negative impact on the QoS.

Moreover, the existing solutions did not take into account the environment in the simulation model such as if the road is highway or urban. Also, they did not simulate the real propagation and mobility parameters which are affected by various factors such as signal fading, multi-path propagation and obstacles. Therefore, lack of applying models which simulate the real environment affects the performance metrics.
In addition, existing security solutions did not consider the malfunction behavior or communication condition when making a decision about malicious node. 
Finally, edge computing was applied on IoT-devices to allow data to be processed near to where it is created rather than crossing long route to reach central server/cloud. Thus, it reduces the delay that is resulted from the packets transmission. Edge computing has been applied on V2X communication which is called vehicular edge computing because it works efficiently with latency-sensitive use cases. This is suitable for situations that require very short latency such as warning messages. However, security and privacy are still serious challenges in vehicular edge computing that require to be considered in the future (\cite{rev1}).


\section{Conclusion}

The rapid evolution of the transportation sector has caused security challenges which made the vehicular network vulnerable to various cyber-attacks that hinder the secure V2X communication. In addition, we should take into account the features of the V2X network while designing the security model.
In this survey paper, we first clarified the key features and architecture of the V2X network. Also, we proposed the threats analysis for V2X enabling technologies. Then, we classified the current security solutions in vehicular networks based on the security method. Also, we presented the comparison and discussion of various security methods. 

Finally, we mentioned the main challenges and future research directions for novel contributions to this research area. As a conclusion, combining cryptography and trust strategies will protect the network from internal and external attacks and thus guarantee high security level.




\section*{References}

 \bibliographystyle{elsarticle-harv} 
 \bibliography{V2X-Ref}


%
%
%
\end{document}